\documentclass[preprintnumbers,amsmath,amssymbm,preprint]{revtex4}
\usepackage{epsfig}
\usepackage{graphicx}
\usepackage{amssymb}

\begin{document}
\title{Numerical evidence for the validity of the inverse hoop conjecture for black holes}
\author{Shahar Hod}
\affiliation{The Ruppin Academic Center, Emeq Hefer 40250, Israel}
\date{\today}

\begin{abstract}
\ \ \ The inverse hoop (IH) conjecture, ${\cal M}_{\text{irr}}\leq {\cal C}_{\text{s}}/4\pi$, 
originally discussed in the context of stationary black-hole configurations, suggests that the surface area of a black hole cannot undergo unconstrained geometric wrinkling (here ${\cal M}_{\text{irr}}=\sqrt{{\cal A}/16\pi}$ 
is the irreducible mass associated with the horizon area ${\cal A}$ of the black hole and ${\cal C}_{\text{s}}$ is 
the circumference length of the smallest ring that can engulf the horizon of the black hole in every direction). 
In the present compact paper we provide evidence supporting the validity of the IH conjecture for non-stationary black-hole configurations. In particular, we show that all the non-stationary black-hole solutions recently constructed by Croft, Garc\'ia-Moreno, and Pani (CGP) [arXiv:2611.13266] satisfy the IH relation.
\end{abstract}
\bigskip
\maketitle

\section{Introduction}

In two-dimensional Euclidean geometry, the isoperimetric theorem \cite{Iso} dictates that the area ${\cal A}$ 
of a connected domain is constrained by its perimeter ${\cal C}$ through the fundamental upper bound
\begin{equation}\label{Eq1}
{\cal A}\leq {{{\cal C}^2}\over{4\pi}}\ ,
\end{equation}
where equality is saturated strictly by a circular boundary. 

Conversely, this standard area-circumference relationship can be violated by warped or crumpled two-dimensional surfaces embedded within a three-dimensional space \cite{Notecir12}. In an analogous manner, the horizon surface area of a $(3+1)$-dimensional black hole can, in principle, scale unboundedly relative to the square of its circumference. 

Nevertheless, it is well-established that black holes in three spatial dimensions frequently exhibit characteristics reminiscent of two-dimensional systems. Most notably, their thermodynamic entropy is governed by the surface area of the event horizon rather than its effective spatial volume \cite{Bekent,Hawent}. Consequently, one might naturally anticipate that, mirroring the classical two-dimensional bound in (\ref{Eq1}), the area of a black hole could be fundamentally restricted by a quadratic function of its circumference.

Motivated by this idea, we have recently conjectured \cite{Hodih} that the surface areas of all stationary $(3+1)$-dimensional black holes 
are bounded from above by the hoop relation
\begin{equation}\label{Eq2}
{\cal A}\leq {{{\cal C}^2_{\text{s}}}\over{\pi}}\ ,
\end{equation}
where ${\cal C}_{\text{s}}$ denotes the circumference of the minimal encircling ring capable of enclosing the black-hole horizon across all azimuthal orientations. 

It is worth recalling that the original hoop conjecture of Thorne \cite{Thorne} asserts that black-hole spacetimes 
are characterized by the simple inequality ${\cal M}\geq{\cal C}_{\text{s}}/4\pi$ \cite{Noterec,Hodst,Peng,Noteunit}. 
However, the precise physical meaning of the mass term ${\cal M}$ in the original hoop relation was not specified in \cite{Thorne}. Note that, using the geometric definition of the irreducible mass of a black hole, 
\begin{equation}\label{Eq3}
{\cal M}_{\text{irr}}\equiv\sqrt{{{\cal A}\over{16\pi}}}\  ,
\end{equation}
the conjectured inequality (\ref{Eq2}) can be expressed in the form
\begin{equation}\label{Eq4}
{\cal M}_{\text{irr}}\leq{{\cal C}_{\text{s}}\over{4\pi}}\  .
\end{equation}
We therefore refer to the conjectured black-hole inequality (\ref{Eq2}) [or, equivalently, the relation (\ref{Eq4})] as the {\it inverse hoop} (IH) conjecture. The proposed IH inequality (\ref{Eq2}), if true, entails that the surface area of a black hole is prevented from exhibiting unconstrained geometric wrinkling \cite{Notenbh}.

Remarkably, it was explicitly shown in \cite{Hodih} that generic Kerr-Newman-(anti)-de Sitter black holes conform to the conjectured area-circumference relation (\ref{Eq2}). What is more intriguing is that the numerical data presented by East \cite{East,Notepri} for the gravitational collapse of non-spinning highly non-spherical collisionless matter configurations indicate that non-stationary (dynamically formed) black holes may also respect the proposed IH bound (\ref{Eq2}).

Most recently, Croft, Garc\'ia-Moreno, and Pani (CGP) \cite{CGP} studied the (in)validity of various formulations of the 
original hoop conjecture \cite{Thorne,Noterec,Hodst,Peng} for axisymmetric families of initial data
describing rotating dust configurations. The main goal of the present compact paper is to test the validity of the 
conjectured IH relation (\ref{Eq2}) for the non-stationary rotating black-hole configurations studied in \cite{CGP}. 

\section{Testing the inverse hoop conjecture with rotating axisymmetric dust configurations}

In the present section we shall test, using the recently published interesting 
numerical data of \cite{CGP}, the validity of the conjectured IH relation (\ref{Eq2}) for non-stationary rotating black-hole configurations. 
To this end, we define the dimensionless physical parameter \cite{NoteHH}
\begin{equation}\label{Eq5}
{\cal H}\equiv {{\pi{\cal A}}\over{{\cal C}^2_{\text{s}}}}\  .
\end{equation}
The conjectured IH relation (\ref{Eq2}) suggests that $(3+1)$-dimensional black holes are characterized by 
the dimensionless inequality  
\begin{equation}\label{Eq6}
{\cal H}\leq1\  .
\end{equation}

In our analysis of the non-stationary initial data sets constructed by CGP \cite{CGP}, we examine three distinct classes of axisymmetric configurations representing rotating dust clouds: spheroids, oblate ellipsoids (pancakes), and prolate ellipsoids (baguettes). 
These configurations are governed by several key parameters with distinct physical meanings: $\rho_0$ denotes the central density parameter controlling the compaction of the matter distribution; $e \equiv \sqrt{1 - [\min(a,b)/\max(a,b)]^2}$ defines the geometric eccentricity \cite{Noteab}, quantifying the deviation of the spatial distribution from spherical symmetry (where $e = 0$ corresponds to a sphere, $a < b$ to oblate pancakes, and $a > b$ to prolate baguettes); and $v_0$ serves as the rotation parameter governing either rigid ($v_N^\phi = v_0 = \text{const}$) or differential ($v_N^\phi(r,z) = v_0(b/[r+\epsilon])[1+\kappa P(\xi;\sigma)]$ \cite{Noterz,Note3p}) angular-velocity profiles that impart rotational energy and angular momentum to the dust fluid.

Below we present a test of the inverse hoop conjecture across all black-hole-forming initial data 
configurations reported in \cite{CGP}. Interestingly, the data presented in Tables I-VIII explicitly demonstrate that all 
nontrivial (nonspherical) black-hole configurations constructed in \cite{CGP} are 
characterized by the relation ${\cal H}\leq1$ and therefore respect the IH conjecture (\ref{Eq2}).

\begin{table}[h!]
\centering
\begin{tabular}{|l|c|c|c|c|c|}
\hline
~ $\rho_0$ ~ & ~ 5.1 ~ & ~ 5.2 ~ & ~ 5.3 ~ & ~ 5.4 ~ & ~ 5.5 ~ \\ \hline
~ ${\cal H}$ ~ & ~ 0.7758 ~ & ~ 0.7952 ~ & ~ 0.8094 ~ & ~ 0.8209 ~ & ~ 0.8309 ~ \\
\hline
\end{tabular}
\caption{The inverse hoop parameter ${\cal H}$ for oblate ellipsoids of fixed eccentricity $e = 0.99$ with varying densities (Table 2 of \cite{CGP}).}
\end{table}

\begin{table}[h!]
\centering
\begin{tabular}{|l|c|}
\hline
~ $\rho_0$ ~ & ~ 7.0 ~ \\ \hline
~ ${\cal H}$ ~ & ~ 0.7814 ~ \\
\hline
\end{tabular}
\caption{The inverse hoop parameter ${\cal H}$ for a prolate ellipsoid of fixed eccentricity $e = 0.99$ (Table 3 of \cite{CGP}).}
\end{table}

\begin{table}[h!]
\centering
\begin{tabular}{|l|c|c|c|}
\hline
~ $e$ ~ & ~ 0.93 ~ & ~ 0.94 ~ & ~ 0.95 ~ \\ \hline
~ ${\cal H}$ ~ & ~ 0.8572 ~ & ~ 0.8422 ~ & ~ 0.8200 ~ \\
\hline
\end{tabular}
\caption{The inverse hoop parameter ${\cal H}$ for oblate ellipsoids of fixed density $\rho_0 = 4.0$ and varying eccentricity $e$ (Table 4 of \cite{CGP}).}
\end{table}

\begin{table}[h!]
\centering
\begin{tabular}{|l|c|c|c|c|}
\hline
~ $e$ ~ & ~ 0.93 ~ & ~ 0.94 ~ & ~ 0.95 ~ & ~ 0.96 ~ \\ \hline
~ ${\cal H}$ ~ & ~ 0.9152 ~ & ~ 0.9030 ~ & ~ 0.8863 ~ & ~ 0.8614 ~ \\
\hline
\end{tabular}
\caption{The inverse hoop parameter ${\cal H}$ for prolate ellipsoids of fixed density $\rho_0 = 5.0$ and varying eccentricity $e$ (Table 5 of \cite{CGP}).}
\end{table}

\begin{table}[h!]
\centering
\begin{tabular}{|l|c|c|c|c|}
\hline
~ $v_0$ ~ & ~ 0.0 ~ & ~ 1.0 ~ & ~ 2.0 ~ & ~ 3.0 ~ \\ \hline
~ ${\cal H}$ ~ & ~ 1.0000 ~ & ~ 0.9998 ~ & ~ 0.9994 ~ & ~ 0.9986 ~ \\
\hline
\end{tabular}
\caption{The inverse hoop parameter ${\cal H}$ for rigidly rotating spheres of fixed central density $\rho_0 = 3.1$ with varying rotation parameter $v_0$ (Table 6 of \cite{CGP}).}
\end{table}

\begin{table}[h!]
\centering
\begin{tabular}{|l|c|c|c|c|}
\hline
~ $v_0$ ~ & ~ 0.0 ~ & ~ 0.5 ~ & ~ 1.0 ~ & ~ 1.5 ~ \\ \hline
~ ${\cal H}$ ~ & ~ 0.7757 ~ & ~ 0.7741 ~ & ~ 0.7687 ~ & ~ 0.7569 ~ \\
\hline
\end{tabular}
\caption{The inverse hoop parameter ${\cal H}$ for rigidly rotating oblate ellipsoids of fixed eccentricity $e = 0.99$ and central density $\rho_0 = 5.1$ with varying rotation parameter $v_0$ (Table 7 of \cite{CGP}).}
\end{table}

\begin{table}[h!]
\centering
\begin{tabular}{|l|c|c|c|c|}
\hline
~ $v_0$ ~ & ~ 0.0 ~ & ~ 4.0 ~ & ~ 8.0 ~ & ~ 12.0 ~ \\ \hline
~ ${\cal H}$ ~ & ~ 0.7813 ~ & ~ 0.7812 ~ & ~ 0.7808 ~ & ~ 0.7800 ~ \\
\hline
\end{tabular}
\caption{The inverse hoop parameter ${\cal H}$ for rigidly rotating prolate ellipsoids of fixed eccentricity $e = 0.99$ and central density $\rho_0 = 7.0$ with varying rotation parameter $v_0$ (Table 8 of \cite{CGP}).}
\end{table}

\begin{table}[h!]
\centering
\begin{tabular}{|l|c|}
\hline
~ $v_0$ ~ & ~ 10.0 ~ \\ \hline
~ ${\cal H}$ ~ & ~ 0.893 ~ \\
\hline
\end{tabular}
\caption{The inverse hoop parameter ${\cal H}$ for a differentially rotating oblate spheroid with parameters $\rho_0 = 10.0$, $v_0 = 10.0$ and $e = 0.99$ (Table 10 of \cite{CGP}).}
\end{table}

\section{Summary and discussion}

The inverse hoop (IH) conjecture suggests the existence of an 
upper bound on the irreducible mass of a black hole which 
is expressed in terms of the circumference of the minimal encircling ring that can engulf the horizon from every 
direction, ${\cal M}_{\text{irr}} \leq {\cal C}_{\text{s}}/4\pi$. 
Equivalently, this inequality limits the horizon surface area ${\cal A}$ in terms of the squared circumference of the minimal encircling ring [see Eq. (\ref{Eq2})]:
\begin{equation}\label{Eq7}
{\cal A}\leq {{{\cal C}^2_{\text{s}}}/\pi}\  .
\end{equation}
If true, the geometric IH relation (\ref{Eq7}) dictates that the horizon of a $(3+1)$-dimensional black hole cannot undergo unconstrained crumpling or wrinkling \cite{Notenbh}. This upper area-circumference bound was originally formulated and 
tested for stationary Kerr-Newman-(anti)-de Sitter black holes \cite{Hodih}. Determining whether it persists in non-stationary regimes remains an important open question in black-hole physics.

In the present compact paper we have tested the validity of the IH conjecture using the numerical data recently published by Croft, García-Moreno, and Pani \cite{CGP} for non-stationary axisymmetric initial data sets describing rotating dust configurations. By computing the dimensionless inverse hoop parameter ${\cal H} \equiv \pi{\cal A}/{\cal C}^2_{\text{s}}$ across a wide parameter space — encompassing rigidly and differentially rotating spheroids, eccentric oblate ellipsoids (pancakes), and prolate ellipsoids (baguettes) characterized by various central densities $\rho_0$, eccentricities $e$, and rotation 
parameters $v_0$ — we have explicitly demonstrated that all the black-hole-forming configurations constructed by CGP \cite{CGP} obey the dimensionless inequality ${\cal H} \leq 1$ (see the data presented in Tables I–VIII).

Taken together with earlier evidence from the gravitational collapse of non-spinning collisionless matter configurations \cite{East}, our results (which are based on the numerical data presented in \cite{CGP}) provide compelling numerical evidence for the general validity of the inverse hoop conjecture beyond stationary equilibrium spacetimes. While a rigorous analytical proof remains an important open challenge, our findings explicitly demonstrate that, even in the presence of significant rotational distortion and extreme deviations from spherical symmetry, black-hole apparent horizons remain strictly constrained by the geometric bound (7).

\bigskip
\noindent {\bf ACKNOWLEDGMENTS}

This research is supported by the Carmel Science Foundation. I would like to thank 
Robin Croft, Gerardo García-Moreno, and Paolo Pani for interesting correspondence. 
I would also like to thank Yael Oren, Arbel M. Ongo, Ayelet B. Lata, and Alona B.
Tea for stimulating discussions.


\begin{thebibliography}{99}
\bibitem{Notecir12} For a two-dimensional domain situated in a three-dimensional manifold, the geometric scale $\mathcal{C}$ is defined by the smallest possible circumference among all circular contours enclosing the region from every perspective.
\bibitem{Noteunit} We use natural units in which $G=c=1$.
\bibitem{Notenbh} Note that the proposed inverse hoop relation (\ref{Eq2}) is restricted in applicability 
to black-hole horizons. Ordinary horizonless bodies can readily be constructed to contravene inequality (\ref{Eq2}); for instance, a heavily cratered celestial body resembling a moon, or alternatively a virus-like particle exhibiting a dense array of surface spikes, will easily violate the area-circumference bound (\ref{Eq2}).
\bibitem{NoteHH} Note that our parameter ${\cal H}$ is related to the dimensionless equatorial and polar 
parameters $\tilde{\cal R}^{\text{AH}}_{\text{P}}$ and $\tilde{\cal R}^{\text{AH}}_{\text{E}}$ of \cite{CGP} 
by the relation 
${\cal H}=\big(\text{max}\{\tilde{\cal R}^{\text{AH}}_{\text{P}},\tilde{\cal R}^{\text{AH}}_{\text{E}}\}\big)^{-2}$ 
[see Eq. (\ref{Eq5}) and Eqs. (4.6)-(4.7) of \cite{CGP}].
\bibitem{Note3p} Here $\xi$ is a dimensionless ellipsoidal coordinate parameter defined such that the interior region 
of an ellipsoid with semi-axes $a$ and $b$ corresponds to $\xi^2 < 1$ \cite{CGP}; 
$P(\xi; \sigma)$ is a smooth $C^\infty$ bump function that determines the spatial distribution profile of the matter density \cite{CGP}; $\sigma$ is a dimensionless control parameter governing the steepness of the density decay in the bump function and the differential rotation profile \cite{CGP}; and $\kappa$ is a dimensionless parameter that regulates the amplitude of the angular velocity variation within the dust cloud \cite{CGP}.
\end{thebibliography}
\end{document}